\documentclass[aps,prl,superscriptaddress,showpacs,twocolumn]{revtex4}
\begin{document}
\title{Competing Interactions, the Renormalization Group and the Isotropic-Nematic Phase Transition}
\author{Daniel G.\ Barci}
\affiliation{Departamento de F{\'\i}sica Te\'orica,
Universidade do Estado do Rio de Janeiro, Rua S\~ao Francisco Xavier 524, 20550-013,  Rio de Janeiro, RJ, Brazil.}
\altaffiliation{Research Associate of the Abdus Salam International Centre for Theoretical Physics}
\author{Daniel A.\ Stariolo}
\affiliation{Departamento de F{\'\i}sica, Universidade Federal do Rio Grande do Sul, CP 15051, 91501-970, Porto Alegre, Brazil} 
\altaffiliation{Research Associate of the Abdus Salam International Centre for Theoretical Physics}

\begin{abstract}
We discuss $2D$ systems with Ising symmetry and competing interactions at different scales. In the framework of
the Renormalization Group, we study the effect of relevant quartic interactions.
In addition to the usual constant interaction term, we analyze the effect of quadrupole interactions in the self consistent Hartree approximation. We show that in the case of 
repulsive quadrupole interaction, there is a first order phase transition to a stripe phase in agreement with the well 
known Brazovskii result. However, in the case of attractive quadrupole interactions there is an isotropic-nematic second 
order transition with higher critical temperature. 
\end{abstract}
%
%
%
\maketitle
The interest in phases with complex translational and/or orientational order is growing up in the condensed matter 
community in the last years. Systems with very different microscopic origin present phase transitions from  
disorder isotropic, homogeneous phases to anisotropic and/or inhomogeneous ones. The physical origin of this behavior is in  many cases associated with the competition between short ranged attractive and long ranged repulsive interactions
\cite{SeAn1995}.
The attractive part tends to form ordered phases or condensates, or in the case of conserved order parameter, tends 
to produce phase separation. However, a long ranged repulsion frustrates this tendency, favoring the emergence of complex phases that break translational and/or rotational symmetry.  
Examples of these systems go from highly correlated quantum systems like quantum Hall samples~\cite{qhs1,qhs2} and High 
$T_c$ superconductors~\cite{htc} to classical systems like ferromagnetic films~\cite{AbKaPoSa1995,VaStMaPiPoPe2000}, 
diblock copolymers~\cite{copolymers} and liquid crystals\cite{lc1,lc2}, to cite but a few. 
All these systems have a special  regime where physical properties are dominated by an extended region in momentum space with large degeneracy. For instance, in Fermi liquids this degeneracy is related with the existence of a Fermi 
surface at low temperatures. In classical systems, the competing interactions lead to a shift of the dominant wave 
vector in the structure factor to a non-zero value. 
In nearly isotropic systems, the low energy degrees of freedom  are kinematically constrained to a thin spherical shell 
of radius $k_0$ determined by interactions.
Examples of scalar and vector order parameters behaving this way were studied by Brazovskii in a seminal 
work~\cite{Brazovskii}. He showed that in systems with a spectrum of fluctuations dominated by a non-zero wave vector,
there is a first order phase transition at a finite temperature form an isotropic to a stripe phase, 
induced by field fluctuations. 
His prediction was experimentally confirmed in the microphase separation transition in diblock 
copolymers~\cite{copolymersexp}. 
It is also observed in Monte Carlo simulations of ultrathin magnetic films with perpendicular
anisotropy~\cite{CaStTa2004} and in the Coulomb frustrated ferromagnet~\cite{GrTaVi2001}, where the first order 
transition is clearly seen. However, theoretical and numerical work on ferromagnetic films
~\cite{AbKaPoSa1995,CaMiStTa2006} and the classic KTHNY theory of two dimensional melting~\cite{KTHNY}, predict a 
more complex phase diagram. For instance, it could be possible to melt the stripes into  a nematic phase, where the 
translational order is lost, but orientational order remains. 

With this motivation we review in this letter the effective low energy theory for systems with competing interactions 
under the perspective of the Renormalization Group (RG)\cite{wilson}.  The presence of a new scale $k_0$ with a 
large momentum space allowed for fluctuations, completely changes the properties of the Gaussian 
fixed point. We found that, upon expanding the angular momentum content of the interaction, the $d=2$ system 
is characterized by an infinite number of relevant quartic coupling constants. This result resembles the Fermi liquid theory 
where there is an infinite number of marginal Landau parameters controlling the fixed point~\cite{FLfixedpoint}. 
In particular, we found that the first nontrivial coupling after the usual local $\phi^4$ theory, represents an
interaction between local quadrupole moments. If this interaction is repulsive, the Brazovskii analysis remains the 
correct one, predicting a first order transition to an inhomogeneous state. However, if this interaction is attractive, 
there is a new instability describing a second order isotropic-nematic phase transition with critical temperature higher 
than the melting transition of the Brazovskii model. We have characterized this $d=2$ nematic critical point in the self 
consistent Hartree approximation. We have computed the critical temperature $T_c$ as well as the critical exponents 
$\beta=1/2$, $\gamma=1$. The conditions for the existence of the nematic phase  and its critical properties are the main results of this letter. 
In the rest of the paper we sketch the analysis  leading to these results. 

In general, the low temperature physics of $2d$ models with competing interactions and Ising symmetry 
is  well described by a coarse-grained Hamiltonian of the type:
\begin{equation}
H_0= \int_{\Lambda} \frac{d^{2}k}{(2\pi)^2}\;\phi(\vec k)\left(r_0 + \frac{1}{m}(k-k_0)^2+ \ldots\right) \phi(-\vec k)
\label{H0}
\end{equation}
where $r_0(T)\sim (T-T_c)$, $k=|\vec k|$ and $k_0=|\vec k_0|$ is a constant given by the nature of the competing 
interactions. $\int_{\Lambda} d^{2}k \equiv  \int_0^{2\pi} d\theta \int_{k_0-\Lambda}^{k_0+\Lambda} dk\;k$
and $\Lambda \sim \sqrt{m r_0}$ is a cut-off where the expansion of the free energy up to quadratic order in the momentum makes 
sense. The ``mass'' $m$ measures the curvature of the dispersion relation around the minimum $k_0$ and  the ``$\ldots$'' 
in eq.(\ref{H0}) means higher order terms in $(k-k_0)$. 
The structure factor has a maximum at $k=k_0$ with a correlation length $\xi\sim 1/\sqrt{m r_0}$. Therefore, near 
criticality ($r_0\to 0$) the physics is dominated by an annulus in momentum space with momenta $k \sim k_0$ and width 
$2\Lambda$.    
This situation is quite similar with fermionic systems at low temperature, where the role of $k_0$ is the Fermi 
momentum, and the reduction of phase space to a spherical shell  centered at the Fermi momentum is 
ruled by the Pauli exclusion principle. In our case of interest, the microscopic physics is very different, but the 
effects of kinematical constraints on  momentum  space are equivalent. 

We would like to determine what kind of interaction terms are relevant to study the low energy physics of a system given 
by the Hamiltonian of eq.(\ref{H0}). The method is similar to the RG for Fermi liquids developed in ref.~\cite{shankar} 
and already applied to the Brazovskii model in ref.~\cite{Hohenberg} to study the first-order transition to the stripe
phase. The standard procedure is to identify a scale transformation that leaves the Gaussian theory invariant, 
and then study the relevance of interactions in the scaling limit very near the circle $k=k_0$. With this aim, as usual, 
we define a small radial wave vector $q=k-k_0$, we reduce the cutoff to $\Lambda/s$, with $s>1$, and  integrate 
over rapid modes $\Lambda/s<|q|<\Lambda$, then we rescale the fields and wave vectors to reestablish the same scale 
and compare the couplings. 
First, note that the kinetic term in eq. (\ref{H0}) is invariant under the rescalings
\begin{eqnarray}
\Lambda' &=& \Lambda/s,  \label{scaling1} \\
q'&=& q s, \label{scaling2}\\
\phi'(q')&=&s^{-3/2} \label{scaling3}\phi(q'/s).
\end{eqnarray} 
Some important comments are in order. These transformations are independent of the dimension of the momentum space. 
This result is very different for systems without competing interactions where $k_0=0$. In this case, the fields would 
scale as $\phi'(q')=s^{-(d+2)/2} \phi(q'/s)$, where $d$ is the spatial dimension of the system. This fact leads to a 
completely different analysis in the case of competing interactions. In fact, the usual concepts of upper and lower 
critical dimensions will change in our case, due essentially to the degeneracy of the lowest energy manifold. 
With this scaling in mind we immediately conclude that the term $r_0 |\phi|^2$ in eq. (\ref{H0}) is relevant as it should 
be, as this term controls criticality.
Let us now analyze a generic quartic interaction (we don't analyze cubic terms in this article since we are interested 
in systems with Ising symmetry $\phi\to -\phi$):
\begin{widetext}
\begin{equation}
H_{\rm int}=\int_\Lambda \frac{d^2k_1}{(2\pi)^2}\frac{d^2k_2}{(2\pi)^2}\frac{d^2k_3}{(2\pi)^2}\frac{d^2k_4}{(2\pi)^2}   
\; u(\vec k_1,\vec k_2,\vec k_3,\vec k_4)\; \phi(\vec k_1)\phi(\vec k_2)\phi(\vec k_3)\phi(\vec k_4) 
\; \delta(\vec k_1+\vec k_2+\vec k_3+\vec k_4).
\label{H4}
\end{equation} 
\end{widetext}
Changing variables $q_i=k_i-k_{0 i} \mbox {,with~}i=1\ldots 4$,  and rescaling the momenta and fields following  eqs. 
(\ref{scaling1}), (\ref{scaling2}) and (\ref{scaling3}), we obtain at ``tree level''
$u'(q'_1,q'_2,q'_3,q'_4)=s^3 u(q'_1/s,q'_2/s,q'_3/s,q'_4/s)$\cite{shiwa}.
We immediately see that the constant term is 
relevant.
 Therefore, at this level of approximation it is enough  to keep the quartic term replacing $q_i=0$ in the expression for 
$u(q_i)$. 
It is  important to note a difference with the case $k_0=0$; even though the coupling $u$ is a constant, in the 
sense that it does not depend on $q$, it still depends on the angles  $\theta_i$ of each $\vec k_{0 i}$, and then 
$u\equiv u(\theta_1,\theta_2,\theta_3,\theta_4)$.
However, these angles are strongly constrained by kinematics. Momentum conservation allows us to eliminate one of them 
in terms of the other three. Furthermore, as the momenta are constrained to move in a very narrow circular region of 
radius $k_0$ and width $\Lambda$, we can fix only two of them and the other two are automatically slaved.
Finally, since the system is rotational invariant, the couplings  can only depend on the difference between these two 
angles, $u(\theta_1,\theta_2,\theta_1+\pi,\theta_2+\pi)=u(\theta_1-\theta_2)=u(\theta)$.
Thus, the quartic interaction is represented, not by  few constants, but by a {\em continuous function of an angle}.
There is still a last constraint, due to the fact that $u(\vec k_1,\vec k_2,\vec k_3,\vec k_4)$ in  Eq. (\ref{H4}) 
is invariant under 
any permutation of the four indexes, implying that $u(\theta)=u(\theta+\pi)$. Therefore, we can expand $u(\theta)$
in a Fourier series, obtaining an infinite set of coupling constants representing angular momentum ``channels''
 of the interaction:
\begin{equation}
u(\theta)=u_0+ u_2\;\cos(2\theta)+u_4\;\cos(4\theta)+\ldots
\label{fourier}
\end{equation}
The first term $u_0$ leads to the usual $\phi^4$ theory considered by Brazovskii in his model for the isotropic-smectic 
transition. In ref.\ \cite{Hohenberg} a detailed RG analysis showed the evolution of the running coupling constants $u_0$ 
and $r_0$ far from criticality, $T<< T_c$. This analysis is justified because it is in this region that the 
fluctuation induced first order transition occurs. In this letter we are interested in another regime, $T\sim T_c$, where 
the whole set of couplings $\{u_0, u_2, u_4, \ldots\}$ have to be considered, in principle, at the same level. 
A detailed analysis of the RG flow will be presented elsewhere
~\cite{inpreparation}.

For simplicity, let us consider the effect of the first two terms. A convenient way of representing the Hamiltonian 
in terms of $u_0$ and $u_2$ in real space is
\begin{equation}
H_{\rm int}=\int d^2x\;\; \left\{  u_0\; \phi^4(\vec{x}) + u_2\; tr\ \hat Q ^2\right\},
\label{HN}
\end{equation}
where 
\begin{equation}
\hat Q_{ij}(\vec{x})= \phi(\vec{x}) \left(\nabla_i\nabla_j-\frac{1}{2}\nabla^2\delta_{ij}\right)\phi(\vec{x})
\label{orderparameter}
\end{equation}
can be easily recognized as the quadrupole moment of the density $\phi^2$~\cite{DanielNematic}.

$\hat Q_{ij}$ is a traceless symmetric tensor, being a natural local order parameter for a phase with orientational
order. In fact, a phase with $\langle\phi(\vec{x})\rangle = 0$, and  $\tilde Q_{ij}\equiv\int dx \langle\hat 
Q_{ij}(\vec{x})\rangle\neq 0$ is a homogeneous phase, with orientational order with the nematic 
symmetry $\theta\to\theta+\pi$.  In the following, we will show that under certain conditions, this is the ordered phase 
produced at the onset of the instability $r_0\sim 0$.

To study this phase transition, we make a self consistent Hartree approximation and analyze the Hamiltonian 
eq. (\ref{HN}) in the same lines of Brazovskii's work~\cite{Brazovskii}. This approximation is exact in an $O(N)$ model 
in the limit $N\to \infty$. In our case, it will present $1/N$ corrections.
In this analysis, we should also consider corrections of the type $\gamma\, tr\hat Q ^4$ with $\gamma>0$. 
Although this term do not enter the critical properties of the system at this level of approximation, it will be important to stabilize the low temperature phase. 

As usual, we replace in eq. (\ref{HN}) 
$\phi^4\to \phi^2 <\phi^2>$ and 
$ tr\ \hat Q ^2\to tr\ \left\{ \phi \left(\nabla_i\nabla_j-\frac{1}{2}\nabla^2\delta_{ij}\right)\phi \right\}
\langle\hat Q_{ij}(\vec{x})\rangle$, where the mean values will be determined self consistently.
With this procedure we obtain a quadratic Hamiltonian in the Hartree approximation given in momentum space by
\begin{equation}
H_{\rm Hartree}=\frac{1}{2}\int \frac{d^2k}{(2\pi)^2}\,
\phi(\vec{k})\left(\beta^{-1}C^{-1}(\vec{k})\right)  \phi(-\vec{k}),
\end{equation}
where the static structure factor $C(\vec{k})$ is now given by
\begin{equation}
C(\vec{k})=\frac{T}{r + \frac{1}{m}(k-k_0)^2-\alpha k^2\cos(2\theta)( u_2+\gamma \alpha^2) },
\label{C}
\end{equation}
and
\begin{equation}
r=r_0+  u_0\int \frac{d^2k}{(2\pi)^2}\;\; C(\vec{k}),
\label{rint}
\end{equation}
where we have chosen $\tilde Q_{ij}=\alpha \left(\hat n_i \hat n_j-\frac{1}{2}\delta_{i,j} \right)$ and we have absorbed unimportant numerical factors in the definition of $u_0$.  $\theta$ is the angle subtended by $\vec k$ with the director $\hat n$. 
Note that the new quartic term explicitly introduces an anisotropy in the structure factor of eq. (\ref{C}). The
coupling has exactly the form of the second term in eq. (\ref{fourier}), as it should be. This
new term is also responsible for a shift in the value of the dominant wave vector.
From the definition of the nematic order parameter, eq. (\ref{orderparameter}), we find that the amplitude of 
$\tilde Q_{ij}$ is given by
\begin{equation}
\alpha=- \frac{1}{2} \int \frac{d^2k}{(2\pi)^2}\;  k^2 \cos(2\theta)\; C(\vec{k})
\label{alphaint}
\end{equation}
which, together with (\ref{C}) and (\ref{rint}) completes a set of equations to be solved self-consistently.

After the exact integration over the angles,  we have integrated  the radial variable $k$ in the limit 
where the correlation length $\xi\sim 1/(\sqrt{m r})$ is much larger than the typical wave-length of the system $1/k_0$.   
It can also  easily be checked that $\alpha=0$ is always a solution of eq. (\ref{alphaint}) and we expect that at high 
temperatures this is the only possible solution at finite $r$. Therefore, upon the $k$ integrations at leading order in $\sqrt{m r}/k_0$, the result can be expanded in powers of $\alpha^2/r$.
It is convenient to write the equations in terms of the adimensional parameters $r\to  r T$, $k_0\to k_0\sqrt{mT}$ 
and $r_0\to \tau= c(1-T_c/T)$,
obtaining at leading order in $(\alpha^2/r)$,  
\begin{equation}
r=\tau+ u_0 k_0 m \frac{1}{\sqrt{r}}+ O\left(\frac{\alpha^2}{r}\right)
\label{r}
\end{equation}  
and
\begin{equation}
\alpha^2\left\{ a_1(r,T)+ a_2(r,T) \frac{\alpha^2}{r} + O\left(\left(\frac{\alpha^2}{r}\right)^2\right) \right\}=0\;,
\label{alpha}
\end{equation}  
where $a_1(r,T)=u_2(8+u_2k_0^5m^3T^2/2r^{3/2})$ and $a_2(r,T)=8r\gamma+k_0^5m^3u_2(\gamma+\frac{15}{64}m^2k_0^4u_2/r^2)/\sqrt{r}$. 

From this result, it is clear that if $u_2>0$, corresponding to repulsive quadrupolar interaction, $a_1(r,T)>0$ and $a_2(r,T)>0$ for all $r$ and $T$. Therefore, in this case,  the only possible solution is $\alpha=0$. The model then reduces to 
that of Brazovskii, and a careful study of eq. (\ref{r}) follows the same lines of ref.\ \cite{Brazovskii}. In this case 
nothing happens at $T=T_c$. The system shows a first order phase transition to a stripe phase, with a melting 
temperature $T_m\sim T_c/(1+r_c) < T_c$ where $r_c\equiv r(\tau=0)= (u_0 k_0 m)^{2/3}$.
However, in the case of attractive quadrupole interactions, $u_2<0$, eqs. (\ref{r}) and (\ref{alpha}) have non 
trivial solutions. For high temperatures $T>T_c$ the only possible solution is $\alpha=0$ as anticipated. This represents 
a high temperature disordered homogeneous and isotropic phase. On the other hand, if $T<T_c$, a nematic phase emerges 
continuously with $\alpha\sim |T-T_c|^{1/2}$.  From eqs. (\ref{r}) and (\ref{alpha}) and the condition $a_1(r_c,T_c)=0$, we can read the critical temperature  to be $T_c= 4/(m k_0^2)\; \sqrt{u_0/u_2}$. 

We have also computed the nematic susceptibility, by coupling  the system to a small external field conjugate to the 
nematic order parameter. Considering for simplicity the  orientation of the field in the same direction than the order 
parameter, we find for the nematic susceptibility $\chi_n\sim 1/(T-T_c)$.
This confirms the second order nature of this transition, with critical exponents $\beta=1/2$ and $\gamma=1$.  
One has to bare in mind that these critical exponents will certainly be modified be fluctuations. The reason is that in our Hartree approximation the order parameter $Q_{ij}$ depends on $\phi$ fluctuations. In other words, it is a Hartree approximation for the order parameter $\phi$, however a mean field in $Q_{ij}$. We expect that upon improving this approximation, the isotropic-nematic transition should be of the Kosterlitz-Thoules type\cite{inpreparation}.

Summarizing, we have analyzed a general model of a scalar field theory with competing interactions in two dimensions. 
This model is dominated by a small circular shell in momentum space, profoundly modifying the critical properties of the 
corresponding model without competition ($k_0=0$). In particular, we have shown that in the framework of the Renormalization
Group, the gaussian fixed point is affected by an infinite set of relevant coupling constants that codify the angular momentum
content of a particular microscopic interaction. We have analyzed the simplest model with two coupling constants, 
corresponding to consider quadrupole moment interactions. We have found that the repulsive case reduces to the well
known Brazovskii model, with a fluctuation induced first order phase transition to a non-homogeneous stripe state. 
However, for attractive quadrupole interactions the phase diagram changes considerably. By applying a self-consistent
Hartree approximation in the fields we found a  phase 
transition from a high temperature disordered phase  to a nematic phase at lower temperatures. The critical temperature 
is higher than the melting temperature found by Brazovskii, opening the possibility of having a more complex phase 
diagram, with an intermediate nematic phase and a possible first order nematic-stripe phase transition.  
Whether $u_2$ is positive or negative in a real system depends on the microscopic interactions. Our results also
highlight how to measure nematic order from data for the structure factor and predict the temperature window between 
$T_m$ and $T_c$ opening the way to look for this phase experimentally, in systems like anisotropic thin film 
ferromagnets and diblock copolymers, among many others.
\acknowledgments
We acknowledge Eduardo Fradkin for useful comments and the 
Abdus Salam International Centre for Theoretical Physics, where part of this work was developed.  
DGB thanks the Complex Fluids Group at UFRGS for their kind hospitality.
The {\em ``Conselho Nacional de Desenvolvimento Cient\'{\i }fico e Tecnol\'{o}gico
CNPq-Brazil''} and the {\em ``Funda{\c{c}}{\~{a}}o de Amparo {\`{a}} Pesquisa do Estado
do Rio de Janeiro''} are acknowledged for the financial support.  
\end{document}